\documentclass[10pt]{article}
\usepackage{graphicx}  
\title{On the Goals of Neutrino Astronomy}
\author{F.~Vissani$^1$  
G.~Pagliaroli$^{1,2}$,
F.L.~Villante$^{1,2}$\\
$^1$ INFN, Laboratori Nazionali del Gran Sasso, Assergi (AQ),
Italy\\
$^{2}$ Universit\`a dell'Aquila, Dipartimento di Fisica, L'Aquila, Italy
}
\begin{document}
\maketitle

\begin{abstract}
What do we mean by neutrino astronomy?
Which  information is it able to provide us and which is its potential?
To address these questions, we discuss three among the most relevant 
sources of neutrinos:
the Sun;
the core collapse supernovae;
the supernova remnants.
For each of these astronomical objects, we describe the state of the art,
we present the expectations and
we outline the most actual problems 
from the point of view of neutrino astronomy.
\end{abstract}



The discovery of the world of elementary particles (including
cosmic rays, gravitons, neutrinos) and the growing interest  in
astrophysical processes has opened the way to new astronomies,
collectively named particle astronomies. In the case of neutrino
astronomy, awarded with the Nobel Prize in Physics in 2002, the
progress was unavoidably related to the solution of the neutrino
puzzles, i.e., the solar neutrino problem and the atmospheric
neutrino anomaly, which pointed out non-standard neutrino
properties. Today, half  a century  after the 
deep theoretical insights of B.~Pontecorvo and the pioneristic detection
of neutrinos from natural sources by the Homestake, KGF and CWI  experiments,
we know a lot about these particles. We
can, thus, go back to the original program and use neutrinos  to
probe the inner parts of the stars, the processes of cosmic ray
acceleration, the primordial Universe, etc..
In this contribution, based on our review paper \cite{todo}, we 
describe the perspectives of neutrino astronomy by discussing
expectations and pending problems  regarding some of the best known 
sources of neutrinos.

\section{Solar neutrinos}
Solar neutrino observations have been planned in the sixties to probe 
the nuclear reactions in the solar core. But this research program has been
hostage,  till very recently, of our ignorance of neutrino
properties. 
Nowadays,
the situation has changed.
 The ``solar" squared mass difference, $\delta m^2$,  is determined at 
the level of about
 $\sim 2\%$, mainly by the KamLAND reactor neutrino experiment.
 The leptonic mixing angle $\theta_{12}$ is fixed at the $\sim 6\%$ 
level essentially by the
 SNO solar neutrino experiment and a robust
 upper limit on the leptonic mixing angle $\theta_{13}$ is obtained
 from the CHOOZ experiment. All this implies that the solar neutrino 
oscillation probability is
 reliably known. Furthermore, there
 is a weak evidence for $\theta_{13}\neq 0$
which is crucial for next steps in neutrino physics
(see the contribution of E.~Lisi for
a complete discussion of our present knowledge of neutrino masses and mixing).

When we consider the Sun as an astrophysical system, the discussion 
acquires more facets and becomes even more interesting.
The solar neutrino experiments have demonstrated that nuclear
reactions 
transforming hydrogen into helium:
\begin{equation}
4 p + 2 e^{-}\rightarrow {}^{4}\!{\rm He} + 2 \nu_e 
\;\;\;\;\;\;\;\;(Q=26.7\ {\rm MeV}),
\end{equation}
 occur inside the Sun. Assuming that the energy radiated
by the Sun on the earth, $K_{\odot}$, is due to nuclear processes
and that the Sun is in equilibrium, the total flux of neutrinos
can be easily estimated as $\Phi_\nu\sim 2 K_{\odot}/ Q\simeq 6.5
\times 10^{10} {\rm cm}^{-2}{\rm s}^{-1}$. However solar neutrinos
have a complicated spectrum resulting from different nuclear
reactions chains (i.e., the PP chains and the CNO cycle)
cooperating for helium production in the Sun. The neutrino
spectrum has to be calculated  by constructing a Standard Solar
Model (SSM), which represents the state-of-the-art theoretical
model of the Sun
 and  provides an important benchmark for all stellar evolutionary 
calculations.

Solar neutrino experiments can check the predictions of the SSMs
which are obtained under a number of hypotheses and are affected
by theoretical uncertainties. At present, the only part of the
solar neutrino spectrum which has been {\em directly} probed  by
experiments concerns some secondary branches of the main chain,
namely the Boron neutrinos (SNO, Super-Kamiokande) and the
Beryllium neutrinos (Borexino). The Boron neutrino flux is
strongly dependent on the central temperature of the Sun,  $\delta
\varphi_{\mbox{\tiny B}}\sim 20 \times \delta T_c$, so that it
allows us to constrain $T_c$ better than 1\%.

The 90\% of the solar neutrino flux is not directly measured. This
concerns the most abundant component, i.e., PP neutrinos; we know
it {\em only if we assume} that the Sun is in a state of
equilibrium. We would like to measure the less abundant but more
energetics PEP neutrinos, since the efficiency of PP and PEP
neutrino producing reactions are strictly related.
%
Also, we did not probe yet the amount of CNO neutrinos. These are
subdominant in the Sun, but extremely important for stars in more
advanced evolutionary stages. CNO neutrinos will also provide us
with a handle to address the problem of the observed photospheric
abundances of elements, that disagree with the predictions and
suggest that  certain theoretical hypotheses of the solar model
need revision. Borexino should be soon able to provide these
important measurements, as discussed in the contribution of
D.~Franco; the main issue is not the signal (there are several
events per day) but rather a sufficient understanding of the
background in the signal region.

All this witnesses that the study of solar neutrinos is a very
lively topic, both on the experimental and on the theoretical
point of view.
Also, solar neutrino detectors
will permit us to study the geoneutrinos.
Borexino is in a leading position also in this respect;
KamLAND and SNO+ will be soon in the position to contribute.

\section{Neutrinos from core collapse supernovae}


Stars with sufficiently large masses, at the end of their life,
are characterized by an iron core inert to nuclear reactions,
surrounded by outer shells in which lighter elements are burned.
The mass of the iron core grows as the stellar evolution proceeds.
When it exceeds the mass of $\sim 1.4~M_\odot$, the core
collapses. This eventually originates a compact stellar object,
most commonly a neutron star and, through mechanisms which still
have to be clarified, produces the explosion of the stellar
mantle, leading to an optical supernova (SN). 
The order of magnitude of the potential energy of the neutron star can 
be estimated through its mass, its radius and the Newton constant $G_N$:
\begin{equation}
{\mathcal E}\sim G_N 
\frac{M_{\mbox{\tiny NS}}^2}{R_{\mbox{\tiny NS}}}=
3.5 \times 10^{53}\left(\frac{M_{\mbox{\tiny NS}}}{1.4\,
  M_{\odot}}\right)^{\!2}
\left(\frac{15 \mbox{ km}}{R_{\mbox{\tiny NS}}}\right)\mbox{erg},
\end{equation}
This is remarkably large, the rest mass being $2.5\times 10^{54}$ erg if 
$M_{\mbox{\tiny NS}}=1.4\, M_\odot$. Most of this energy 
is released in neutrinos of all types.

However, the frequency of core collapse SN in the Milky
Way is very low. In fact the expected number of SN, the number of
the progenitors (i.e., the stars with masses $> 8 M_\odot$) and of
their descendants of a certain age, such as pulsars and supernova
remnants (SNR), are linked among them from the relation:
\begin{equation}
\frac{4\ 10^5 \mbox{ stars}}{2\ 10^7 \mbox{ years}}\sim \frac{1
\mbox{ SN}}{50 \mbox{ years}}\sim \frac{4\ 10^4 \mbox{
pulsars}}{2\ 10^6\mbox{ years}}\sim\frac{20 \mbox{ SNR}}
{10^3\mbox{ years}}.
\end{equation}
that assumes that all these populations, 
that are short-lived on galactic time-scale, are in equilibrium among them
(more discussion in the proceedings of IFAE 2005, held in Catania).
We close this introduction by recalling that due to galactic dust, only 1 SN out of 7-8 of them has
been seen; the last one in 1604 by Keplero.

In 1987, a supernova was seen in the Large Magellanic Cloud and
four neutrino detectors announced evidences of signals, possibly
attributable to  electron antineutrinos emitted by the supernova.
It should be kept in mind that, in conventional detectors, the
inverse beta decay of electron antineutrinos on free protons,
$\bar\nu_e p \to e^+ n$, is the most important nuclear reaction to
reveal events of 10-20 MeV of energies.
Despite many puzzling aspects of these observations, the common
view is that the 20-30 events seen by Kamiokande-II, IMB and
Baksan agree with the expectations. In particular, in the
assumption that the electron antineutrinos are 1 sixth of the
emitted neutrino signal, the data seem to confirm the total
emitted energy corresponding to the formation of a neutron star.
The 5 events seen by the fourth detector, LSD, occurred several
hours before the others cannot be accommodated in the simplest
scenario for neutrino emission. They require more complicated
models with several emission phases, that are being developed and
that we will not consider in the following.

In the standard scenario (known as neutrino assisted-or
delayed-explosion or Bethe and Wilson scenario)
the main phases that give observable neutrino fluxes are:\\
{\em 1) The rapid accretion around the nascent neutron star.}
The positrons yield observable antineutrinos
by interacting with the nucleons via
 $e^+ n\to p \bar\nu_e$:
\begin{equation}
L_{accr}\sim N_n\sigma_{e^+ n}T_a^4\sim 5\times 10^{52}\ 
\frac{\mbox{erg}}{\mbox{sec}}\left(\frac{M_a}{0.1\ 
M_\odot}\right)\left(\frac{Y_n}{0.6} \right)
\left(\frac{T_a}{2\ \mbox{MeV}}\right)^{\!6} \label{cinque}
\label{laccr}
\end{equation}
Here,  $T_a$ and $M_a$ are the temperature of the plasma (of the
positrons) and the mass of neutrons exposed to the flux of
positrons; $Y_n$ is the neutron fraction. This phase of emission
lasts a fraction of a second, precedes the
shock  revival and it is crucial for the explosion to occur.\\
{\em 2) The thermal cooling of the protoneutron star.} This happens
on a time scale of about 10 seconds. The luminosity is
proportional to the radiating area:
\begin{equation}
L_{cool}\sim R_c^2\; T_c^4\sim 5\times 10^{51}\
\frac{\mbox{erg}}{\mbox{sec}}\left(\frac{R_c}{10 \mbox{
km}}\right)^{\!2}\left(\frac{T_c}{5\mbox{ MeV}}\right)^{\!4} \label{sei}
\end{equation}
Here, $T_c$ is the initial temperature of the emitted antineutrinos
and $R_c$ is the radius of the sphere that characterizes the emission,
that is expected to have the size of the neutron star. On comparing
with the luminosity in eq.~\ref{cinque}, it is evident
that the thermal phase is much less luminous than the first one.

Recent analyses of the data 
performed using a two phases emission model
showed an evidence for the first luminous phase of about 2.5
sigma.
This result is particularly interesting in view of the difficulty to simulate the 
explosion with computers.
It is curious that this result was obtained so late; but this can be understood considering   
that a large part of the recent discussion was guided by particle physics considerations (mostly,  on neutrino oscillations) rather than astrophysical considerations.


%


It is pretty evident that we will learn a lot from a future
galactic supernova. Consider, e.g., that Super-Kamiokande would
detect about $10^5$ events from a supernova exploding in the
location of the Crab Nebula (2 kpc),  allowing us to probe the
details of the neutrino emission. Moreover, it will become
possible to do astronomy with neutrinos {\em only}: in fact, the
elastic scattering events as seen in a \v{C}erenkov type detector
will permit us to reconstruct the direction of an event in the Galactic
center with a precision of a few degrees. Also, the analysis of
$\bar\nu_e$ permits us to predict with 10 ms precision the moment
of maximum crunch (when the matter bounces on the core of the
neutron star) which is plausibly associated to a strong emission
of gravitational waves.

In summary,  the importance of collecting neutrinos from a
galactic supernova should not be underestimated. The fact that
core collapses are rare in the Milky Way in human standards
should not let us forget that neutrinos will offer us an unique
chance to progress on the understanding of these extraordinary
events. More discussion in the appendix.


%

\section{Neutrinos from supernova remnants}
The Ginzburg and Syrovatskii conjecture, formulated in 1964,
expresses the fact that the energy stored in galactic cosmic rays
is one order of magnitude smaller than the kinetic energy of the
supernova remnants (SNR) of the Milky Way:
\begin{equation}
\frac{\rho_{\mbox{\tiny CR}} V_{\mbox{\tiny 
CR}}}{\tau_{\mbox{\tiny CR}}}\approx 0.1 \times \frac{{\cal
L}_{\mbox{\tiny 
SN}}}{\tau_{\mbox{\tiny SN}}}
\end{equation}
This suggests the idea that SNR act as cosmic ray accelerators.
The most plausible mechanism to realize this possibility
was proposed by Fermi in 1949
and is known as ``diffusive shock wave acceleration''.
We do not have yet a complete understanding of this phenomenon,
but the above conjecture is widely considered of great appeal.

A few SNR were recently observed to radiate gamma well above the TeV.
This could be explained if the cosmic rays, accelerated
in the SNR,
collide with the surrounding medium and produce copious fluxes of mesons,
that eventually decay and lead to gammas, e.g., $\pi^0\to \gamma\gamma$.
It is not yet possible to exclude that (part of) the observed
radiation is produced by electromagnetic processes. In
order to reach a definitive proof of the
hadronic origin of the observed gammas
more detailed studies are needed.

In the assumption that the observed radiation is  produced from hadronic processes,
one expects that a flux of high energy neutrinos is also emitted by SNR.
In hadronic processes, in fact, charged mesons are abundantly produced and their decay
yields charged leptons and neutrinos.
Since the background due to atmospheric neutrinos drops rapidly with
energy, 
there is some hope that these neutrinos can be observed
by the modern neutrino telescopes.
 It is not difficult to calculate the neutrino fluxes expected from
 SNR 
with known gamma-ray spectra.
Since the relative amount of charged and neutral mesons produced in 
hadronic interactions is almost fixed, and
since the fluxes of gamma and neutrinos are not attenuated by the 
diffuse medium in which they propagate,
the gamma and neutrino fluxes are linked by a simple linear relation:
\begin{equation}
F_{\nu_\mu}(E_\nu)+F_{\bar{\nu}_\mu}(E_\nu)=\\
0.66 F_\gamma\!\left(\frac{E_\nu}{1\!-\!r_{\!\pi}}\right) \!+\! 0.02 
F_\gamma\!\left(\frac{E_\nu}{1\!-\!r_{\!K}}\right)\!+
\int_0^1 \!  {\kappa}(x) F_\gamma\!\left(\frac{E_\nu}{x}\right)\! dx
\end{equation} \label{flo}
where $r_{\pi,K}=(m_\mu/m_{\pi,K})^2$ and where:
\begin{equation}
\kappa(x)=
\left\{
\begin{array}{ll}
x^2 (33.8-54.3 x)  & \mbox{ if }x<r_K \\
(1-x)^2 (-0.63+12.45 x)  & \mbox{ if }x>r_\pi \\
0.04+0.20 x +7.44 x^2-7.53 x^3 & \mbox{ otherwise}
\end{array} \label{kel}
\right.
\end{equation}
The above formula includes the contribution of charged pions and 
charged kaons to neutrino production and
the effects of neutrino oscillations.
It is essentially model independent and does not require any 
parameterization of the gamma ray flux $F_\gamma$.
We can thus apply it directly to the gamma ray observational 
data to calculate the neutrino fluxes (and their uncertainties)
and  the corresponding signal in neutrino telescopes.

\begin{figure}[t]
\centerline{\includegraphics[width=0.20\textwidth,angle=270]{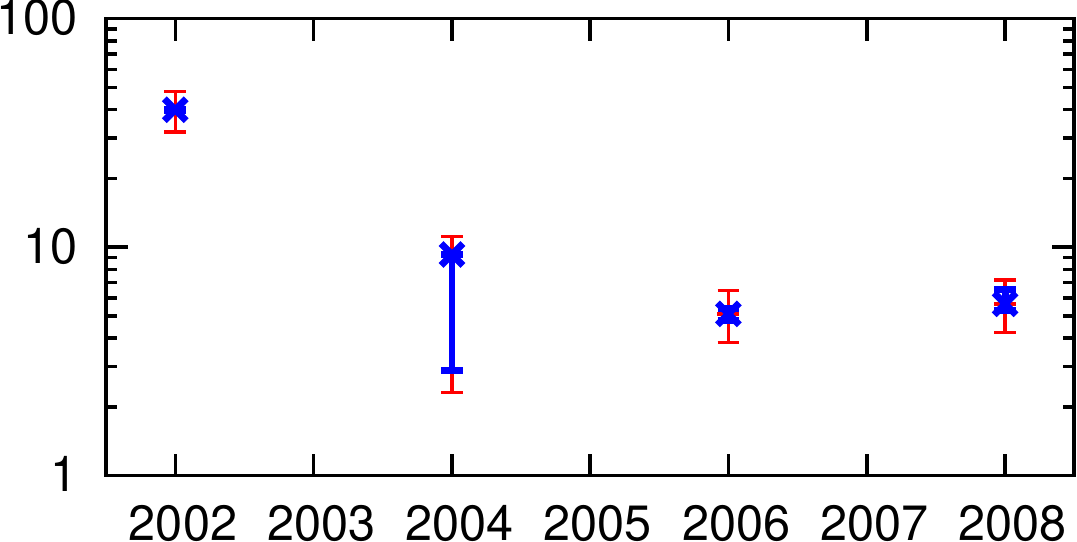}}
\vskip1mm
\caption{\em\small Predictions for the SNR  RX J1713.7-3946.
In the abscissa, year of prediction; in the ordinate, number of muon events
per km$^2$ per year above a threshold of 50 GeV.
In blue, the expected number of events and their error due to the 
uncertainties on the VHE gamma rays spectrum; in red, a 
conservative estimation of the theoretical error, set to 20\%.}
\label{fig1}
\vskip-5mm
\end{figure}

In Fig.~\ref{fig1}, we summarize the expected signal for the best studied SNR: RX J1713.7-3946; the last three points are consistent among them and their difference reflects the improved knowledge of the gamma ray flux.
For this SNR, one expects about 2.5 signal events on top of 1
background event per year and above a threshold of 1 TeV in an ideal
detector of 1 km$^2$ of area located in the Mediterranean. There is
the hope that the number of events from another nearby  SNR, Vela Jr,
is a few times larger. We are waiting to know whether the $\gamma$-ray
observations from H.E.S.S.\ will support these hopes.

\section{Summary}
We close remarking the following three major points:\\
{\em 1.} There is a lively research program on solar neutrinos. 
Borexino is providing the main results;
interesting spinoff measurements such as reactor and geo-$\bar\nu_e$ 
are also possible.\\
{\em 2.} At present, the processes happening in a 
gravitational collapse are only partly understood.
The main possibility to study them is given just by
future neutrino observations;
the paucity of galactic supernovae
does not diminish the importance of this studies.\\
{\em 3.} The neutrinos from supernova remnants could be the turning
point to test the origin of the galactic cosmic rays; they could be
observable with an exposure of several km$^2\times$y.\\

These discussions are in different stages of maturity, but it is
important to emphasize that they concern the same discipline,
neutrino astronomy, that involves astrophysicists, particle
physicists and nuclear physicists and that sees Italy in a leading
role.

\subsection*{Acknowledgments}
F.V.\ thanks the Organizers of IFAE09 for stupendous hospitality and 
F.~Ferroni for the public discussion that lead us to 
develop the considerations listed in the following appendix.

\appendix\section{A limit on the occurrence of galactic supernovae}
As mentioned above, it is not possible to use the astronomical records 
to know the absolute rate of core collapse SN $f$ in the Milky
Way, due to dust extinction of the emitted light. The only firm information 
comes from the fact that in about 25 years of observations, no
neutrino telescopes happened to record an event; applying Poisson statistics, this amounts to the 
90\% C.L.\ upper limit  
$f<1\mbox{ SN}/11\mbox{ y}$.

The problem of dust extinction is less severe for other galaxies, 
that we see head-on. If, additionally, we are able to reliably know 
(or to put a limit on) the {\em relative} rate of occurrence of 
supernovae in another galaxy that we observe, we could use the 
astronomical records to get information on the absolute frequency 
in the Milky Way.

E.g., suppose that the rate of core collapse SN is 3 times lower in 
Andromeda, due to different galaxy mass and stellar population, 
which seems a conservative statement and that can be refined. 
Since the last supernova has been observed in 1885,
we can apply again Poisson statistic to get the 90\% limit on the 
absolute frequency in Andromeda
$f_A<1\mbox{ SN}/54\mbox{ y}$. 
This implies the following limit on the occurrence of galactic supernovae:
\begin{equation}
f<1\mbox{ SN}/18\mbox{ y},
\end{equation}
that is significantly more stringent than the existing direct limit. 
This limit becomes more (resp., less) tight 
presuming that the supernova in 1885 was not due to
 a core collapse event
(resp., that some 
light extintion could be present for Andromeda).

The above discussion shows an interesting direct link between the ordinary 
astronomy and neutrino astronomy. Our argument is close in spirit to 
the one invoked to infer the value 
$f= 1\mbox{ SN}/(30-70)\mbox{ y}$ in the Milky Way, based on the 
observation of supernovae in other galaxies and on the correlation 
of this frequency and the observed (blue) luminosity of the galaxies.  
As for the previous argument, one needs to correlate  the properties 
of the observed galaxies to those of the Milky Way to make the point.


\begin{thebibliography}{0}
\bibitem{todo} G. Pagliaroli, F.L. Villante, F. Vissani,
``Neutrini Dallo Spazio'',
{\em Nuovo Saggiatore 25}, no. 3-4 (2009) 5-19
[freely available online on the SIF web site].
\end{thebibliography}
\end{document}